\documentstyle[aps,psfig]{revtex}
\newcommand{\be}{\begin{equation}}
\newcommand{\ee}{\end{equation}}
\newcommand{\ben}{\begin{eqnarray}}
\newcommand{\een}{\end{eqnarray}}
\newcommand{\ov}{\overline}
\begin{document} 

 
\title{Complete Factorization of Equations of Motion in
Supersymmetric Field Theories}
\author{D. Bazeia,$^a$ J. Menezes,$^a$ and M. M. Santos$^b$}
\address{$^a$Departamento de F\'\i sica, Universidade Federal
da Para\'\i ba, Caixa Postal 5008, 58051-970 Jo\~ao Pessoa PB,
Brazil\\$^b$Departamento de Matem\'atica,
Universidade Estadual de Campinas, Caixa Postal 6065, 13081-970 Campinas SP,
Brazil}

\maketitle

\begin{abstract}

\vspace{12pt}

We investigate bosonic sectors of supersymmetric field theories. We consider
superpotentials described by one and by two real scalar fields, and we show
how the equations of motion can be factorized into a family of first order
Bogomol'nyi equations, so that all the topological defects are of the
Bogomol'nyi-Prasad-Sommerfield type. We examine explicit models, that engender
the $Z_N$ symmetry, and we identify all the topological sectors, illustrating
their integrability.
\vspace{18pt}

\centerline{PACS numbers: 11.27.+d, 11.30.Pb, 11.30.Er}

\vspace{18pt}
\end{abstract}
\vskip1pt

Domain walls are defect structures that appear in diverse branches of physics.
They usually live in three spatial dimensions as bidimensional objects that
arise in systems described by potentials that contain at least two isolated
degenerate minima. They involve energy scales as different as the ones that
appear in Condensed Matter \cite{cm} and in Cosmology \cite{cos}.

A lot of attention has been drawn recently to domain walls in field theories,
in models that have been investigated under several distinct motivations
\cite{fmv90,ato91,cqr91,sg,wit,95,97,dsh97,gto99,bbr00,bve00,nol00,bms01}.
A very specific motivation concerns the presence of
domain walls arising in between non zero vacuum expectation values of scalar
fields in supergravity \cite{cqr91,sg}. Another line deals with the formation
of defects inside domain walls \cite{wit,95,97}. A great deal of attention has
also been drawn to $SU(N)$ supersymmetric gluodynamics, where nonperturbative
effects give rise to gluino condensates that may form according to a set
of $N$ isolated degenerate chirally asymmetric vacua, from where domain
walls spring interpolating between pairs of vacua \cite{dsh97}.

The interest in domain walls in general widens because of the interplay
between Field Theory and the low energy world volume dynamics of branes
in String Theory \cite{sb,sb1,sb2,sbf}. In the case of intersection of
defects, in particular in Ref.~{\cite{bbr00}} some aspects of wall junctions
have been investigated when the discrete symmetry is the $Z_N$ symmetry,
in systems described by a single complex field, with the superpotential
\be
W(\varphi)=\varphi-\frac{1}{N+1}\varphi^{(N+1)},\;\;\;N=2,3,...
\ee
The second work in Ref.~{\cite{bbr00}} has shown that the tensions
of the topological defects that appear in these systems can be cast
to the form
\be
\label{wN}
t_{N,k}=\frac{2N}{(N+1)}\,\sin\left(\frac{k\pi}{N}\right),\;\;\;\;
k=1,...,\biggl[\frac{N}{2}\biggr]
\ee
where $[N/2]$ is the biggest integer not bigger than
$N/2$ itself. In Ref.~{\cite{fmv90}} the same result was obtained, in an
investigation concerned with exact integrability, due to the presence
of infinitely many conserved currents -- see also Ref.~{\cite{dka99}}
for other investigations, in particular on the explicit form
of the BPS solutions at large N.

In the present work we examine the bosonic portion of supersymmetric
theories, similar to the above models, that are of general interest to
supersymmetry. Our investigations follow the lines of the former work
\cite{bms01}, and bring new results on the presence of domain defects
in models described by one and two real scalar fields. We start investigating
systems with a single real scalar field. We write the Lagrangian density in
the standard form
\be
{\cal L}=\frac12\,\partial_\alpha\phi\,\partial^\alpha\phi-\frac12\,W^2_{\phi}
\ee
$W=W(\phi)$ is the superpotential, and $W_\phi=dW/d\phi$.
We search for defect structures, for static solutions of the equation
of motion. We suppose the static solutions depend only on $x$, on a single
spatial coordinate. The equation of motion becomes
$d^2\phi/dx^2=W_{\phi}W_{\phi\phi}$. We examine the energy of the static
configurations, and we associate with the second order ordinary
differential equation of motion the two first order ordinary differential
equations $d\phi/dx=\pm W_{\phi}$. These first order equations are
Bogomol'nyi equations and their solutions are named BPS states, which
saturate the lower bound in energy.

We see that solutions of the first order equations solve the second order
equation of motion, and the proof follows by direct differentiation of the
first order equations. We also see that one associates with the second order
equation two first order equations. This fact suggests the possibility of
factorizing the equation of motion into first order Bogomol'nyi equations.
This is indeed the case, and we demonstrate this property by examining the
ratio $R(\phi)=(d\phi/dx)/(dW/d\phi)$. We use this expression to write
\be
\frac{dR(\phi)}{dx}=\Biggl[W^2_{\phi}-\left(\frac{d\phi}{dx}\right)^2\Biggr]
\frac{W_{\phi\phi}}{W^2_{\phi}}
\ee
We notice that $dW^2_{\phi}/dx=2(d\phi/dx)W_{\phi}W_{\phi\phi}$ and that 
\be
\frac{d}{dx}\left(\frac{d\phi}{dx}\right)^2=
2\,\frac{d\phi}{dx}\,W_{\phi}\,W_{\phi\phi}
\ee
where we have used the equation of motion. These results show that the
quantity $S(\phi)=W^2_{\phi}-(d\phi/dx)^2$ does not depend on $x$
when $\phi(x)$ solves the equation of motion. For those $\phi(x)$ we have
that $\lim_{x\to-\infty}\phi(x)\to{v^k}$, with ${v^k}$ being a vacuum state,
and $\lim_{x\to-\infty}(d\phi/dx)=0$. Also, $\lim_{x\to-\infty} W_{\phi}=0$,
since the vacuum states are extrema of the superpotential. Thus we get that
$S(\phi)$ vanishes, and this allows writing $R(\phi)=\pm1$, which gives the
first order Bogomol'nyi equations. This result shows that the second order
equation of motion is completely equivalent to the two first order Bogomol'nyi
equations. A direct consequence of this result is that in such models all the
topological solutions are of the BPS type, that is, these models do not admit
the presence of non-BPS states. The first order Bogomol'nyi equations can be
readily integrated, so our result offers a new way of showing exact
integrability.

We illustrate our result with some examples. There are several models of
systems of a single real scalar field. We work with dimensionless quantities,
and we consider the superpotential
\be
\label{wphi}
W^n(\phi)=\frac1{n+1}\,\phi^{n+1}-\frac1{n+3}\,\phi^{n+3},\;\;\;n=0,1,...
\ee
Widely known examples are described by $n=0,1$, and identify the $\phi^4$
and the $\phi^6$ models, respectively. These models engender the $Z_2$
symmetry, and for $n=0$ there are two asymmetric vacua, $v^1=-1$ and $v^2=1$.
For $n=1,2,...$ there are three vacua, the former two and another one,
symmetric, $v_0=0$, that appear with multiplicity $n$. For $n=0$ the first
order equations give the standard defects, $\phi(x)=\pm\tanh(x)$. These BPS
solutions connect the two distinct asymmetric vacua. For $n=1$ the solutions
are such that $\phi^2(x)=(1/2)[1\pm\tanh(x)]$. These BPS states connect the
symmetric vacuum $v_0=0$ to the asymmetric ones, $v^{1,2}=\pm1$. The models
defined by the above superpotential are exactly solvable; the solutions can
be known implicitly for $n=2,3,...$ The tensions are: for $n=0$, $t^0=4/3$,
and for $n=1,2,...$, $t^n=2/(n+1)(n+3)$. They show that the width of the
defect increases for increasing $n$.

The main difficulty to generalize the former result is that it is unclear
how to write something like the ratio $R(\phi)$ if we deal with two or
more fields. However, we have found an interesting possibility, that appears
in the case of two real scalar fields, when the superpotential is harmonic on
the two fields. The issue here is that for $W(\phi,\chi)$ harmonic we
can work with a complex superpotential, written as $W(\varphi)$, in terms
of the complex field $\varphi=\phi+i\chi$ -- see Ref.~{\cite{bbr00}}.
In this case the ratio $R(\phi)$ generalizes to $R(\varphi)$, in the form
\be
\label{rc}
R(\varphi)=\frac{1}{\ov{W'({\varphi})}}\,\frac{d\varphi}{dx}
\ee
The superpotential $W({\varphi})$ is holomorphic, and we consider
systems that are described by the Lagrangian density
\be
{\cal L}=\frac12\,\partial_{\alpha}{\ov\varphi}\,\partial^{\alpha}\varphi-
\frac12|W'(\varphi)|^2
\ee
where $W'(\varphi)=\partial W/\partial\varphi$. The specific form of the
potential shows that the critical values of the superpotential are minima
of the potential. We search for defect structures, for static solutions of
the equation of motion. We suppose the static solutions depend only on $x$,
so the equation of motion becomes
\be
\label{soceq}
\frac{d^2\varphi}{dx^2}=W'(\varphi)\,{\ov{W''(\varphi)}}
\ee
We examine the energy of the static configurations to show that the second
order ordinary differential equation is associated with a family
of first order ordinary differential equations. They are
\be
\label{foceq}
\frac{d\varphi}{dx}={\ov{W'(\varphi)}}\,e^{-i\xi}
\ee
where $\xi$ is some real constant. These first order equations are Bogomol'nyi
equations. As usual, solutions of these first order equations
also solve the second order equation of motion. The proof goes as follows:
we differentiate the first order Eq.~(\ref{foceq}) to get
\ben
\frac{d^2\varphi}{dx^2}&=&\frac{d}{dx}{\ov{W'(\varphi)}}\,e^{-i\xi}\nonumber
\\
&=&{\ov{W''(\varphi)}}\,\frac{d{\ov\varphi}}{dx}\,e^{-i\xi}=
{\ov{W''(\varphi)}}\,W'(\varphi)
\een
In the above equation, in the last equality we have used the first order
Bogomol'nyi Eq.~(\ref{foceq}) once again.

The above models are somehow similar to the former models, described by
a single real scalar field. Thus they may also lead to factorization of
the equations of motion into first order Bogomol'nyi equations. We make
this reasoning unambiguous by concentrating on the ratio $R(\varphi)$
defined by Eq.~(\ref{rc}). We can write
\ben
\frac{dR(\varphi)}{dx}&=&\left(\frac1{\ov{W'(\varphi)}}\right)^2\Biggl[
\,{\ov{W'(\varphi)}}\,\frac{d^2\varphi}{dx^2}-
\frac{d\varphi}{dx}\,{\ov{W''(\varphi)}}\,
\frac{d{\ov\varphi}}{dx}\Biggr]\nonumber
\\
&=&\left(\frac1{{\ov{W'(\varphi)}}}\right)^2\Biggl[
\Bigl|W'(\varphi)\Bigr|^2-\Bigl|\frac{d{\varphi}}{dx}
\Bigr|^2\Biggr]\,{\ov{W''(\varphi)}}
\een
We notice that
\be
\frac{d}{dx}|W'(\varphi)|^2
={\ov{W'(\varphi)}}W''(\varphi)\frac{d\varphi}{dx}
+W'(\varphi){\ov{W''(\varphi)}} \frac{d{\ov\varphi}}{dx}
\label{eq1}
\ee
We also notice that
\be
\frac{d}{dx}\biggl|\frac{d\varphi}{dx}\biggr|^2
=W'(\varphi){\ov{W''(\varphi)}}\,\frac{d{\ov\varphi}}{dx}+
{\ov{W'(\varphi)}}\,W''(\varphi)\frac{d\varphi}{dx}
\label{eq2}
\ee
where we have used the equation of motion.
We use the above Eqs.~(\ref{eq1}) and (\ref{eq2}) to see that now the
quantity $S(\varphi)=|W'(\varphi)|^2-|{d\varphi}/{dx}|^2$ is such that
$dS(\varphi)/dx=0$, that is, it does not depend on $x$ when
$\varphi(x)$ solves the equation of motion. For those $\varphi(x)$ we have
that $\lim_{x\to-\infty}\varphi(x)\to{v^k}$, with ${v^k}$ a vacuum state, and
$\lim_{x\to-\infty}(d\varphi/dx)=0$. Also, $\lim_{x\to-\infty} W'(\varphi)=0$,
because the vacuum states are critical points of the superpotential. Thus we
get that $S(\varphi)$ vanishes, so that $R(\varphi)$ does not depend on $x$.
Also, $|d\varphi/dx|=|W'(\varphi)|$ and this
allows writing $|R(\varphi)|=1$. Thus we get that $R(\varphi)=e^{-i\xi}$
for some constant $\xi$, which leads to the first order Bogomol'nyi
equations.

This result shows that the second order equation
of motion is equivalent to the first order Bogomol'nyi equations. This is
valid with the boundary conditions $\lim_{x\to-\infty}\varphi(x)\to{v^k}$,
$\lim_{x\to-\infty}(d\varphi/dx)=0$, as required by the topological
solutions. A direct consequence of this result is that in such models all
the topological solutions are of the BPS type, that is, these models do not
support non-BPS states. Furthermore, we recall that in systems of real scalar
fields, one may occasionally run into solutions that engender no topological
feature. They are named nontopological solutions, and have been found for
instance in Ref.~{\cite{mrw}}. We first notice that in systems of real scalar
fields described by some superpotential, the nontopological solutions cannot
appear as solutions of the first order Bogomol'nyi equations, because they
should have zero energy, and this is the energy of the vacuum states. We add
this to our former result to obtain another result, that in systems of two
real scalar fields, when the superpotential is harmonic there is no room
for non topological solutions.

To illustrate these results let us consider the superpotential 
\be
\label{sp}
W^n_N(\varphi)=\frac1{n+1}\varphi^{n+1}-\frac1{N+n+1}\varphi^{N+n+1}
\ee
It is defined by $N=2,3,...$ and by $n=0,1,...$ The potential is given by
\be
V=\frac12({\ov\varphi}\varphi)^n
\left(1-{\ov\varphi}^N\right)\left(1-\varphi^N\right)
\ee
The models engender the $Z_N$ symmetry, irrespective of the specific
value of $n$, and the limit $n\to0$ leads us back to the model described by
Eq.~(\ref{wN}). Below we obtain the tension associated with every BPS
solution in a closed form.

The critical values of $W^n_N(\varphi)$ can be readily obtained. For $n=0$
they are $v^k_N=\exp(i\xi^k_N)$, where $\xi^k_N=2\pi(k/N)$ and
$k=1,2,...,N$. For $n\neq0$, we have to add to
this set of extrema the special value $v_0=0$, that appears with multiplicity
$n$. Thus, while in the case $n=0$ we get $N$ asymmetric phases, degenerate,
in the case $n\neq0$ we have to include another phase, that breaks no symmetry
of the original system. This includes the possibility of a chirally symmetric
phase in $SU(N)$ supersymmetric gluodynamics. 

We investigate the tension of the defect structures. Since the defects
solve first order Bogomol'nyi equations, their tensions are obtained as:
for $n\neq0$, for solutions that connect the origin $v_0$
to any of the minima $v^k_N$ we can write $t^n_{N,0}=|\,W^n_N(1)\,|$, that
define the radial sectors. The result is, for $N$
arbitrary, and for $n=1,2,...$
\be\label{tradial}
t^n_{N,0}=\frac{N}{(n+1)(N+n+1)}
\ee
We set $N=2$ to get $t^n_{2,0}=2/(n+1)(n+3)$, which reproduces former
result, obtained below Eq.~(\ref{wphi}).

The other sectors represent defects that connect vacua in the unit circle.
Here we have to calculate the quantity $|W^n_N(v^N_N)-W^n_N(v^k_N)|$.
The result can be cast to the form
\be
\label{tcircle}
t^n_{N,k}=2\,t^n_{N,0}\;\biggl|\sin\biggl[(n+1)\frac{k\pi}{N}\biggr]\biggr|
\ee
where $k=1,2,...,[N/2]$ classifies the minima on the circle that the defect
connects: $k=1$ for first neighbours, $k=2$ for second neighbours, and so
forth. We set $n=0$ to get back to the former result, Eq.~(\ref{wN}). 

We use the superpotential (\ref{sp}) to write the family of Bogomol'nyi
equations in the form, in terms of the real fields $\phi$ and $\chi$, 
\ben
\label{r21}
\frac{d\phi}{dx}&=&{\bf P}^n_N \cos\xi +{\bf Q}^n_N\sin\xi
\\
\label{r22}
\frac{d\chi}{dx}&=&{\bf Q}^n_N\cos\xi -{\bf P}^n_N\sin\xi
\een
Here we have introduced the quantities
\be
{\bf P}^n_N=P_n(1-P_N)+Q_nQ_N
\ee
and
\be
{\bf Q}^n_N=Q_n(1-P_N)-P_nQ_N
\ee
which are defined in terms of the functions
\be
\label{sp}
P_k={\cal R}e\,(\phi-i\chi)^k,\;\;\;\;\;
Q_k={\cal I}m\,(\phi-i\chi)^k
\ee

We illustrate the general situation considering
particular values of $N$ and $n$. The case $n=0$ is special, because it
does not admit the origin in configuration space as a vacuum state.
We consider $n=0$ and $N$ even, $N=2j,j=1,2,...\,$.
We examine the Bogomol'nyi equations for solutions
that connect the vacua $(\pm1,0)$, which are farthest neighbours in the
set of $N=2j$ minima. In this case the equations demand that $\chi=0$,
so there are only one-field solutions, obeying $d\phi/dx=\pm(1-\phi^{2j})$.
In the case $j=1$ we get to the standard tanh defects that we have already
described. For $j=2,3,...$ we have other solutions, that can be known
implicitly. They behave as $\phi(x)=x-x^{2j+1}/(2j+1)$ in the limit $x\to0$.
They are similar to the standard defect that we have already found.
However, they are thinner and then more energetic as $j$ increases,
as it can be readily verified. We use the tension result to get
$t^0_{2j,j}=4j/(2j+1)$. This tension increases from $4/3$ for $j=1$
up to the finite value $2$ in the limit $j\to\infty$. There are
other solutions, that connect other vacua in the unit circle.

In particular we consider the case $N=2, n=0$. The superpotential is
$\phi-\phi^3/3+\phi\chi^2$. It is instructive to see this model
with another superpotential, more general,
$W(\phi,\chi)=\phi-\phi^3/3-r\phi\chi^2$. The model with
$N=2, n=0$ is obtained for $r=-1$, the value
that makes the superpotential harmonic. For $r\in (0,1/2)$, however, the
more general model maps the anysotropic $XY$ model \cite{cm}: there are
one-field solutions describing Ising walls, and two-field solutions
describing Bloch walls -- see the last work in Ref.~{\cite{bbr00}}.
For $r=-1$, for non vanishing $\chi$ one must have $\phi^2=1+\chi^2/3$,
so there is no two-field solution connecting the minima $(\pm1,0)$ for $r=-1$.
The lesson we learn here is that the condition $W_{\phi\phi}+W_{\chi\chi}=0$
certainly simplifies the model, favoring its integrability.

Similar behavior is also present in the case $N=2$, for $n$ arbitrary. Here
all the vacuum states are in the $\phi$ axis. The first order Bogomol'nyi
equations {\it only} support topological solutions for $\chi=0$, irrespective
of the value of $n$. The Bogomol'nyi equations effectively reduce to the form
\be
\frac{d\phi}{dx}=\pm\phi^n\left(1-\phi^2\right)
\ee
which supports no nontrivial
two-field solutions. They are the first order equations that appear in the
superpotential of Eq.~(\ref{wphi}). The result is that for $N=2$ and $n$
arbitrary the systems reduce to systems of just one field. The tensions
of the defects for $N=2$ are: for $n=0$, $t^0_{2,1}=4/3$, and for $n\neq0$,
$t^n_{2,0}=2/(n+1)(n+3)$, as noted below Eq.(\ref{tradial}).

We examine $t^n_{N,k}$ to see that it vanishes for some values of $N,n$,
and $k$. The condition $n=N-1$ is special, because now the models
{\it only} support $N$ radial sectors, that describe solutions
connecting $(0,0)$ to any of the $N$ vacua in the unit circle. For
$n=N-1$ the BPS sectors effectively reduce to the sector described
by just one field, and the Bogomol'nyi equations become
\be
\frac{d\phi}{dx}=\pm \phi^{N-1}(1-\phi^N)
\ee
The corresponding tensions are
$t^{N-1}_{N,0}=1/2N$. The case $N=2$ reproduces the case $n=1$ of the
superpotential in Eq.~(\ref{wphi}). Other values of $N$ and $n$ give rise
to radial sectors, and to sectors that connect vacua in the unit circle. An
example of this is given by $N=6$ and $n=2$. This model supports six radial
sectors, with tension $t^2_{6,0}=2/9$ and six other sectors, connecting first
neighbours in the unit circle, with tension $t^2_{6,1}=4/9$, twice the tension
of the radial solutions.

Evidently, the new idea of factorizing equations of motion into
first order Bogomol'nyi equations is of direct interest to
supersymmetry. It can be used in diverse applications, in particular to
investigate the presence of BPS walls interpolating between
distinct pairs of vacua, and exact integrability. The complete factorization
of equations of motion in systems described by $W(\phi,\chi)$ harmonic, or
by $W(\varphi)$ holomorphic, ensures BPS feature to domain walls that spring
in such systems. Because of the associated BPS character, these domain walls
partially preserve the supersymmetry.

The superpotential $W(\phi,\chi)$ in general depends on a set of
parameters, and these parameters can be adjusted to make $W(\phi,\chi)$
harmonic or not. An example of this can be found for instance in the second
paragraph below Eq.~(\ref{sp}), in the model used to map the anisotropic $XY$
model, which supports Ising and Bloch walls -- in the case $r=-1$ the
superpotential there considered becomes harmonic, but the model does not
support Bloch wall anymore. On the other hand, harmonic and non-harmonic
superpotetials of the form $W(\phi,\chi)$ may be used to distinguish two types
of supersymmetry, one with ${\cal N}=1$ and the other, enhanced to the case
${\cal N}=2$. This behavior has been examined for instance in
Ref.~{\cite{rsv01}}. In that work, we can also find Wess-Zumino
models with two complex fields, and there it was shown that the
system supports topological sectors of the non-BPS type. Thus, the
non-BPS solutions can be used as examples that show that our result
does not hold in Wess-Zumino models with two or more complex fields. 

We thank F.A. Brito, J.R. Morris, J.R.S. Nascimento, and R.F. Ribeiro
for discussions, and CAPES, CNPq, PROCAD and PRONEX for partial support.

\end{document}